\documentclass[aip,jmp,reprint,unsortedaddress]{revtex4-1}
\usepackage{amssymb}
\usepackage{amsmath}
\usepackage{graphicx}
\usepackage{txfonts}
\usepackage{amsfonts}

\begin{document}

\title{Dimerization-assisted energy transport in light-harvesting complexes}
\author{S. Yang}
\affiliation{Institute of Theoretical Physics, Chinese Academy of Sciences, Beijing
100190, China}
\author{D. Z. Xu}
\affiliation{Institute of Theoretical Physics, Chinese Academy of Sciences, Beijing
100190, China}
\author{Z. Song}
\affiliation{School of Physics, Nankai University, Tianjin 300071, China}
\author{C. P. Sun}
\email{suncp@itp.ac.cn}
\homepage{http://www.itp.ac.cn/~suncp}
\affiliation{Institute of Theoretical Physics, Chinese Academy of Sciences, Beijing
100190, China}

\begin{abstract}
We study the role of the dimer structure of light-harvesting complex
II (LH2) in excitation transfer from the LH2 (without a reaction
center (RC)) to the LH1 (surrounding the RC), or from the LH2 to
another LH2. The excited and un-excited states of a
bacteriochlorophyll (BChl) are modeled by a quasi-spin. In the
framework of quantum open system theory, we represent the excitation
transfer as the total leakage of the LH2 system and then calculate
the transfer efficiency and average transfer time. For different
initial states with various quantum superposition properties, we
study how the dimerization of the B850 BChl ring can enhance the
transfer efficiency and shorten the average transfer time.
\end{abstract}

\maketitle

\section{Introduction}

To face the present and forthcoming global energy crisis, human
should search for clean and effective energy source. Recently the
investigations on the basic energy science for this purpose has
received great attention and
experienced impressive progress based on the fundamental physics \cite%
{Fleming08,arti}. In photosynthetic process, the structural elegance
and chemical high efficiency of the natural system based on pigment
molecules in transferring the energy of sunlight have stimulated a
purpose driven investigation \cite
{Venturi08,HuXiChe971,Johnson08,Plenio091,Nori09,Guzik09,Nori08,Fleming09,Plenio092,Guzik08,Mukamel09},
finding artificial analogs of porphyrin-based chromophores. These
artificial systems replicate the natural process of photosynthesis
\cite{arti} so that the much higher efficiencies could be gained
than that obtained in the conventional solid systems \cite{arti}. It
is because one of the most attractive features of photosynthesis is
that the light energy can be captured and transported to the
reaction center (RC) within about 100ps and with more than 95\%
efficiency \cite{HuXiChe971,Fleming94}.

Actually, in most of the plants and bacterium, the primary processes
of photosynthesis are almost in common
\cite{Fleming94,HuXiChe972,Venturi08}: Light is harvested by antenna
proteins containing many chromophores; then the electronic
excitations are transferred to the RC sequentially, where
photochemical reactions take place to convert the excitation energy
into chemical energy. Most recent experiments have been able to
exactly determine the time scales of various transfer processes by
the ultra-fast laser technology \cite{exp1,exp2,exp3}. These great
progresses obviously offer us a chance to quantitatively make clear
the underlying physical mechanism of the photosynthesis, so that
people can construct the artificial photosynthesis devices in the
future to reach the photon-energy and photon-electricity conversions
with higher efficiency. For example, quantum interference effects in
energy transfer dynamics \cite{Guzik08} has been studied for the
Fenna-Matthews-Olson (FMO) protein complex, and it was found
\cite{Guzik09} that, for such molecular arrays, the spatial
correlations in the phonon bath and its induced decoherence could
affect on the efficiency of the primary photosynthetic event. The
present paper will similarly study the influences of spatial
structure on the primary processes of photosynthesis for the
light-harvesting complexes II (LH2).

In the past, by making use of the x-ray crystallographic techniques, the
structure of light-harvesting system has been elucidated \cite%
{HuXiChe96,Venturi08}. In the purple photosynthetic bacteria, there exist
roughly two types of light-harvesting complexes, referred to as
light-harvesting complex I (LH1) and light-harvesting complex II (LH2). In
LH1, the RC is surrounded by a B875 bacteriochlorophyll (BChl) ring with
maximum absorption peak at 875 nm. The LH2 complex, however, does not
contain the RC, but can transfer energy excitation to the RC indirectly
through LH1. In the purple bacteria, LH2 is a ring-shaped aggregate built up
by $8$ (or $9$) minimal units, where each unit consists of an $\alpha \beta $%
-heterodimer, three BChls, and one carotenoid. The $\alpha \beta $%
-heterodimers, i.e., $\alpha $-apoproteins and $\beta $-apoproteins
constitute the skeleton of LH2, while the BChls are embedded in the scaffold
to form a double-layered ring structure. The top ring including $16$ (or $18$%
) BChl molecules is named as B850 since it has the lowest-energy
absorption maximum at 850 nm. The bottom ring with 8 BChls is called
B800 because it mainly absorbs light at 800 nm. In every minimal
unit, the carotenoid connects B800 BChl with one of the two B850
BChls. Excitation is transferred
from one pigment to the neighbor one through the F\"{o}ster mechanism \cite%
{HuXiChe971}, while the electron is spatially transferred via the
Marcus mechanism \cite{Leegwater96}. Generally, it is independent of
the global geometry configuration of the system.

In the present paper, we will study the energy transfer procedure in LH2 by
considering the structure dimerization of the B850 ring. It has been
conjectured that the dimerized inter-pigment couplings can cause the energy
gap to protect the collective excitations \cite{HuXiChe972}. Indeed, like
the the Su-Schrieffer-Heeger model for the flexible polyacetylene chain \cite%
{SSH}, the dimerization of the spatial configuration with the Peierls
distorted ground state will minimize the total energy for the phonon plus
electron. As it is well known, this model exhibits a rich variety of
nonlinear phenomena and topological excitations including the topological
protection of the quantum state transfer \cite{Song}. Similarly, we will
show that, when the B850 ring in LH2 is dimerized the excitation transfer
efficiency may be enhanced to some extent.

Based on the open quantum system theory, we simply model the excited
and un-excited states of a BChl pigment as a quasispin. The
excitation transfer is represented by the total leakage from a LH2.
Using the master equation, we calculate the efficiency of excitation
transfer and the average transfer time in low temperature for
various initial states with different superposition properties. The
results explicitly indicate that the dimerization of couplings
indeed enhances the quantum transport efficiency and shortens the
average transfer time.

\begin{figure}[ptb]
\includegraphics[bb=49 366 550 770, width=8 cm, clip]{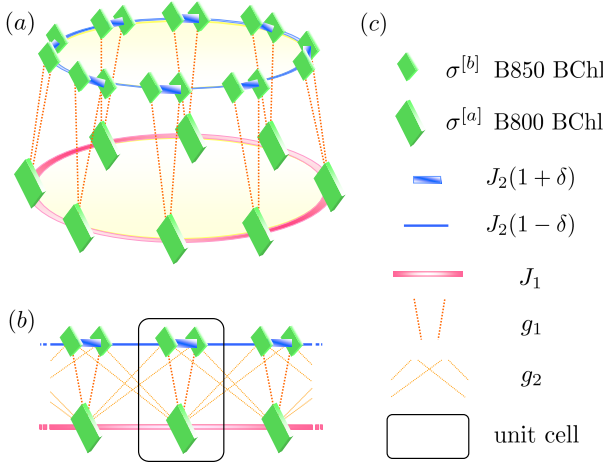}
\caption{(color online) The model setup of the light-harvesting
complex II constructed by 8 unit cells. The couplings between the
neighboring
quasi-spins in the B850 ring is dimerized as $J_{2}(1+\protect\delta)$ and $%
J_{2}(1-\protect\delta)$. $g_{1}$ denote the nearest couplings
between B850 BChls and B800 BChls, while $g_{2}$ denote the next
nearest couplings  between B850 BChls and B800 BChls. (a)
Illustration of the whole system with $g_{2}=0 $. (b) Detailed
drawing of three unit cells and their non-local couplings. (c)
Legends.} \label{model1}
\end{figure}

This paper is organized as follows. In Sec. II, a double-ring XY model with $%
N$ unit cells is presented to simulate the LH2 system. In Sec. III, the
energy transfer process is described by the quantum master equation. The
transfer efficiency $\eta \left( t\right) $ and the average transfer time $%
\tau $ are introduced to characterize the dynamics of the system. In Sec.
IV, we represent the master equation in the momentum space and show that
only the $\left( k,k\right) $-blocks of the density matrix are relevant to
energy transfer. In Sec. V, it is found that the transfer efficiency $\eta
\left( t\right) $ and the average transfer time $\tau $ of an arbitrary
initial state can be obtained through the channel decomposition. Some
numerical analysis of $\eta ^{\left[ A,k\right] }\left( t_{0}\right) $ and $%
\tau ^{\left[ A,k\right] }$ for all the $k$-channels are presented in Sec.
VI. They show that a suitable dimerization of the B850 BChl ring can enhance
the transfer efficiency and shorten the average transfer time. Conclusions
are summarized at the end of the paper. In Appendix A, we provide an
alternative way to deal with the energy leakage problem. In Appendix B,
detail derivations of transforming the master equation from the real space
to the $k$-space are given. The approximate solution of $\tau ^{\left[ A,k%
\right] }$ for $k=0$ and $k=\pm \pi $ channel is shown in Appendix C.

\section{Model setup}

The simplified model of LH2 is\ shown in Fig. \ref{model1}. All the
bacteriochlorophylls (big and small green squares) are modeled by the
two-level systems with excited state $|e_{j}^{[c]}\rangle $, ground state $%
|g_{j}^{[c]}\rangle $,\ and energy level spacing $\Omega _{c}$. The raising
and lowering quasi-spin operators of the $j$th two-level system on the $[c]$%
\ ring is expressed as
\begin{equation}
\sigma _{j}^{+[c]}=|e_{j}^{[c]}\rangle \langle g_{j}^{[c]}|\text{, }\sigma
_{j}^{-\left[ c\right] }=|g_{j}^{[c]}\rangle \langle e_{j}^{[c]}|,
\end{equation}%
where $[c]=$\ $\left[ a\right] $ ($\left[ b\right] $) denotes the B800
(B850) BChl ring. Approximately, all the couplings are supposed to be of XY
type \cite{Johnson08}. This simplification enjoys the main feature of
excitation transfer. \ The Hamiltonians
\begin{equation}
H_{a}=\frac{\Omega _{a}}{2} \sum_{j=1}^{N}\sigma _{j}^{z\left[
a\right] }+J_{1}\sum_{j=1}^{N}\left( \sigma _{j}^{+\left[ a\right]
}\sigma _{j+1}^{-[a]}+\mathrm{H.c.}\right) \label{h1}
\end{equation}%
and
\begin{align}
H_{b}& =\frac{\Omega _{b}}{2} \sum_{j=1}^{2N}\sigma _{j}^{z\left[
b\right] }+J_{2}\sum_{j=1}^{N}[\left( 1+\delta \right) \sigma _{2j-1}^{+%
\left[ b\right] }\sigma _{2j}^{-\left[ b\right] }  \notag \\
& +\left( 1-\delta \right) \sigma _{2j}^{+\left[ b\right] }\sigma _{2j+1}^{-%
\left[ b\right] }+\mathrm{H.c.}]  \label{h2}
\end{align}%
with $N=8$,\ describe the excitations of the B800 and\ B850 BChl rings,\
respectively.\ In the B850 BChl ring, the parameter $\delta \neq 0$
characterizes the dimerization due to the spatial deformation of the
flexible B850 BChl ring in LH2. The coupling constants of $H_{b}$ are
dimerized as $J_{2}\left( 1+\delta \right) $ and $J_{2}\left( 1-\delta
\right) $ since the intra-unit and inter-unit Mg-Mg distance between
neighboring B850 BChls may be different. The non-local XY type interaction

\begin{align}
H_{ab}& =g_{1}\sum_{j=1}^{N}\left[ \sigma _{j}^{+\left[ a\right] }\left(
\sigma _{2j-1}^{-\left[ b\right] }+\sigma _{2j}^{-\left[ b\right] }\right) +%
\mathrm{H.c.}\right]  \notag \\
& +g_{2}\sum_{j=1}^{N}\left[ \sigma _{j}^{+\left[ a\right] }\left( \sigma
_{2j-3}^{-\left[ b\right] }+\sigma _{2j-2}^{-\left[ b\right] }+\sigma
_{2j+1}^{-\left[ b\right] }+\sigma _{2j+2}^{-\left[ b\right] }\right) +%
\mathrm{H.c.}\right]  \label{Hspin}
\end{align}%
is used to describe the interaction between the B800 and B850 BChl rings.

In the single excitation case, the quasi-spin can be represented with a
spinless fermion with the mapping%
\begin{equation}
\sigma _{j}^{+\left[ a\right] }\leftrightarrow A_{j}^{\dag },\sigma
_{2j-1}^{+\left[ b\right] }\leftrightarrow B_{j}^{\dag },\sigma _{2j}^{+%
\left[ b\right] }\leftrightarrow C_{j}^{\dag }
\end{equation}%
from the spin space $V_{s}=C_{2}^{\otimes 3N}$ to the subspace $V_{F}$ of\
the Fermion Fock space spanned by
\begin{equation}
\left\{ \left\vert O,j\right\rangle =O_{j}^{\dag }\left\vert 0\right\rangle
\text{ }|\text{ }O=A,B,C;j=1,2,\cdots ,N\right\} .
\end{equation}%
Hereafter, let us represent the site index as $\left( O,j\right) $, where $j$
refers to a unit cell shown in Fig. \ref{model1}, and $O=A,B,C$ to a
position type inside the unit cell. In the subscripts, the site index $%
\left( O,j\right) $ is written as $Oj$ for simplicity. The vacuum state of
the Fermion system $\left\vert 0\right\rangle $ corresponds to the state
that all the quasi-spins are in their ground states,
\begin{equation}
\left\vert 0\right\rangle \leftrightarrow \prod_{j=1}^{N}\left\vert g_{j}^{
\left[ a\right] }\right\rangle \otimes \prod_{j=1}^{2N}\left\vert g_{j}^{%
\left[ b\right] }\right\rangle .
\end{equation}%
Then the total Hamiltonian $H_{S}=H_{a}+H_{b}+H_{ab}$ of LH2 is mapped into
\begin{align}
H_{S}& =\sum_{j=1}^{N}\left[ \Omega _{a}A_{j}^{\dag }A_{j}+\Omega _{b}\left(
B_{j}^{\dag }B_{j}+C_{j}^{\dag }C_{j}\right) \right]  \notag \\
& +\sum_{j=1}^{N}\left\{ J_{1}A_{j}^{\dag }A_{j+1}+g_{1}A_{j}^{\dag }\left(
B_{j}+C_{j}\right) \right.  \notag \\
& +J_{2}[\left( 1+\delta \right) B_{j}^{\dag }C_{j}+\left( 1-\delta \right)
C_{j}^{\dag }B_{j+1}]  \notag \\
& \left. +g_{2}A_{j}^{\dag }\left( B_{j+1}+B_{j-1}+C_{j+1}+C_{j-1}\right) +%
\mathrm{H.c.}\right\} .  \label{Hparticle}
\end{align}%
In the present work, no multi-fermion interactions are considered for
simplicity.

On the other hand, we use the Holstein-Primakoff transformation \cite%
{HPtransform} to map the quasi-spin into bosons. The excitations of
the BChls can be described by quasi-spins with the total angular
momentum $S$. Then $D=A,B,C$ can be regarded as the annihilation
operators
of bosons for the Fock space spanned by%
\begin{eqnarray}
\{\left( D_{j}^{\dag }\right) ^{n_{D,j}}\left\vert 0\right\rangle \text{ }|%
\text{ }D &=&A,B,C;j=1,\cdots ,N;  \notag \\
n_{D,j} &=&0,1,\cdots ,2S\}.
\end{eqnarray}%
For $S>1/2$, one local bacteriochlorophyll has more than one excited states.
In this case, higher order coherence could be included for further
generalization.

In the following, we focus on the single excitation case. Then the
temperature should be suitable to ensure there is no higher order
excited state.

\section{Transfer efficiency and average transfer time via the master
equation}

\begin{figure}[tbp]
\includegraphics[bb=27 374 563 777, width=8 cm, clip]{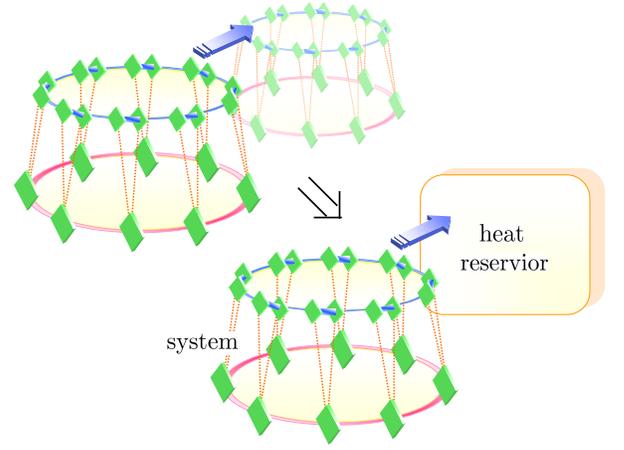}
\caption{(color online) The first excited LH2 is treated as an open system
while the other LHs are regarded as heat reservoirs. The energy transfer
process is equivalent to the excitation leakage from the B850 BChl ring of
the LH2 system to the environment.}
\label{model2}
\end{figure}

Next we consider the energy transfer from an initial state
\begin{equation}
\widehat{\rho }\left( 0\right) =\sum_{j,l}\rho _{Aj,Al}\left( 0\right)
\left\vert A,j\right\rangle \left\langle A,l\right\vert ,
\end{equation}%
which is a coherent superposition or a mixture of those local states $%
\left\vert A,j\right\rangle $ on the B800 ring. As time goes by, the initial
state will evolves a state distributing around both the B800 and the B850
rings. Since there exists a difference of chemical potential. $\Delta \Omega
=\Omega _{a}-\Omega _{b}$, energy is transferred between the two rings
during the time evolution. For an isolated LH2 system, such energy transfer
is coherent, namely, the system oscillates between the B800 and the B850
rings . However, when a LH2 is coupled to a heat reservoir with infinite
degrees of freedom, irreversible energy transfer occurs. As illustrated in
Fig. \ref{model2}, in the real photosynthetic system, energy is transferred
from one LH2 to another LH2 or LH1 through the B850 ring \cite{HuXiChe972}.
Therefore, we regard the first excited LH2 as an open system, and the sum of
others as the a heat reservoir. The energy transfer now can be manipulated
as the energy leakage from the B850 ring to the environment.

In order to describe such a procedure that the excitations are finally
transferred from the B850 ring to the heat reservoir, the Markovian master
equation
\begin{equation}
\frac{d\widehat{\rho }}{dt}=-i\left[ H_{S},\widehat{\rho }\right] +{\mathcal{%
L}}\left( \widehat{\rho }\right)  \label{mastereq}
\end{equation}%
in the Lindblad form is employed for determining the time-evolution of the
density matrix. Here two kinds of loss processes, dissipation and dephasing,
are considered as Lindblad terms
\begin{equation}
{\mathcal{L}}\left( \widehat{\rho }\right) =\sum_{j=1}^{N}\left[ {\mathcal{L}%
}_{\mathrm{diss},j}\left( \widehat{\rho }\right) +{\mathcal{L}}_{\mathrm{deph%
},j}\left( \widehat{\rho }\right) \right] .
\end{equation}%

We suppose that each quasi-spin on the B850 ring is coupled to an
independent heat reservoir \cite{Guzik09}, which reflects the local modes of
phonons and other local fluctuations. Then the dissipation from the $j$th
unit cell is described as
\begin{equation}
{\mathcal{L}}_{\mathrm{diss},j}\left( \widehat{\rho }\right) =\Gamma
_{j}\sum_{O=B,C}(O_{j}\widehat{\rho }O_{j}^{\dag }-\frac{1}{2}\left\{
O_{j}^{\dag }O_{j},\widehat{\rho }\right\} ),  \label{dissipation}
\end{equation}%
where $\left\{ \cdot ,\cdot \right\} $ denotes the anti-commutator.
Here, the sink rate $\Gamma _{j}$ at the $j$th point may be site
dependent. For the dynamics constrained on the subsystem described
by $O$ operators, the last term of
${\mathcal{L}}_{\mathrm{diss},j}\left( \widehat{\rho }\right) $
gives contribution $-\Gamma _{j}\rho _{Oj,Oj}$ to $d\rho
_{Oj,Oj}/dt$, thus dissipation results\ in the reduction of the
total population. Therefore, the dissipation term Eq.
(\ref{dissipation}) represents the incoherent transfer of energy
into the environment.

On the other
hand, the dephasing term reads%
\begin{equation}
{\mathcal{L}}_{\mathrm{deph},j}\left( \widehat{\rho }\right) =\Gamma
_{j}^{\prime }\sum_{O=B,C}(O_{j}^{\dag }O_{j}\widehat{\rho }O_{j}^{\dag
}O_{j}-\frac{1}{2}\left\{ O_{j}^{\dag }O_{j},\widehat{\rho }\right\} ).
\label{dephasing}
\end{equation}%
Compared with the dissipation term, the dephasing one ${\mathcal{L}}_{%
\mathrm{deph},j}\left( \widehat{\rho }\right) $ does not contribute\
to any time local change of the probability distribution, i.e., the
derivative of the diagonal elements of the density matrix is
irrelevant to this term. Thus the total population
$\sum_{O=A,B,C}\sum_{j}\rho _{Oj,Oj}$ would be conserved if only the
dephasing term were present. However, the dephasing process is also
incoherent since it make the nondiagonal elements of the density
matrix tend to zero.

The above two contributions force the LH2 system to ultimately reach a
steady state $\widehat{\rho }_{\mathrm{steady}}=\left\vert 0\right\rangle
\left\langle 0\right\vert =\widehat{\rho }_{v,v}$, namely, in the long-time
limit, all excitations are sinked away. The same steady state is obtained
from Eq. (\ref{mastereq}) in the super-operator form%
\begin{equation}
\frac{d}{dt}[\rho ]=M[\rho ],
\end{equation}%
where $[\rho ]$ denotes the column vector defined by all matrix elements in
some order, and the super-operator $M$ is determined by

\begin{equation}
M[\rho ]=[-i\left[ H_{S},\widehat{\rho }\right] +{\mathcal{L}}\left(
\widehat{\rho }\right) ].
\end{equation}%
In this sense the\ steady state is just the non-trivial eigenstate of $M$
with vanishing eigen-energy. Usually, from $\det M=0,$ the steady state can
be found.

However, we are interested in the system dynamics on a short timescale,
i.e., how soon can the excitations be transferred from one LH2 to the other
light-harvesting complexes? To this end, the transfer efficiency $\eta
\left( t\right) $ is defined as the population $\rho _{v,v}\left( t\right) $
of the vacuum state $\left\vert 0\right\rangle $ at time $t$,
\begin{equation}
\eta \left( t\right) =\rho _{v,v}\left( t\right) .
\end{equation}%
The corresponding master equation (\ref{mastereq})%
\begin{align}
\frac{d\rho _{v,v}}{dt}& =\sum_{j=1}^{N}\Gamma _{j}\left\langle 0\right\vert
\left( B_{j}\widehat{\rho }B_{j}^{\dag }+C_{j}\widehat{\rho }C_{j}^{\dag
}\right) \left\vert 0\right\rangle  \notag \\
& =\sum_{j=1}^{N}\Gamma _{j}\sum_{O=B,C}\rho _{Oj,Oj}
\end{align}%
means that only the first term of ${\mathcal{L}}_{\mathrm{diss},j}\left(
\rho \right) $ contributes to the time derivatives of $\rho _{v,v}\left(
t\right) $. The transfer efficiency is given by the integral of the above
formula \cite{Johnson08,Guzik09,Plenio092},%
\begin{equation}
\eta \left( t\right) =\int_{0}^{t}\sum_{j=1}^{N}\Gamma _{j}\sum_{O=B,C}\rho
_{Oj,Oj}\left( t^{\prime }\right) dt^{\prime }.  \label{Eeta}
\end{equation}%
The average transfer time $\tau $ is further defined as \cite%
{Guzik09,Johnson08}
\begin{eqnarray}
\tau &=&\lim_{t\rightarrow \infty }\frac{1}{\eta \left( t\right) }%
\int_{0}^{t}t^{\prime }\sum_{j=1}^{N}\Gamma _{j}\sum_{O=B,C}\rho
_{Oj,Oj}\left( t^{\prime }\right) dt^{\prime }  \notag \\
&=&\frac{1}{\overline{\eta }}\int_{0}^{\infty }t^{\prime
}\sum_{j=1}^{N}\Gamma _{j}\sum_{O=B,C}\rho _{Oj,Oj}\left( t^{\prime }\right)
dt^{\prime },  \label{Etau}
\end{eqnarray}%
where usually
\begin{equation}
\overline{\eta }=\lim_{t\rightarrow \infty }\eta \left( t\right) =1.
\end{equation}%
Therefore, an efficient energy transfer requires not only a perfect
transmission efficiency $\eta $ but also a short average time $\tau $.

In Appendix A, we present an equivalent non-Hermitian Hamiltonian method,
which can also be utilized to study the dynamics of the open system.

\section{$k$-space representation of the master equation}

In this section we present the $k$-space representation of the above master
equation, so that we can reduce the dynamics of time evolution in some
invariant subspace. If all the dissipation and dephasing rates are
homogeneous on the B850 BChl ring, i.e., $\Gamma _{j}=\Gamma $ and $\Gamma
_{j}^{\prime }=\Gamma ^{\prime }$, the whole system has translational
symmetry. For each unit cell containing three BChls shown in Fig. \ref%
{model1}, we introduce the Fourier transformation,
\begin{equation}
O_{k}^{\dag }=\frac{1}{\sqrt{N}}\sum_{j=1}^{N}e^{ikj}O_{j}^{\dag }
\label{fourier}
\end{equation}%
for $O=A,B,C$. Then in the $k$-space the Hamiltonian (\ref{Hparticle}) is
represented as $H_{S}=\sum_{k}H_{k}$ with%
\begin{align}
H_{k}& =2J_{1}\cos kA_{k}^{\dag }A_{k}+\{\left( g_{1}+2g_{2}\cos k\right)
\left( A_{k}^{\dag }B_{k}+A_{k}^{\dag }C_{k}\right)  \notag \\
& +J_{2}\left[ \left( 1+\delta \right) +\left( 1-\delta \right) e^{-ik}%
\right] B_{k}^{\dag }C_{k}+\mathrm{H.c.}\}.
\end{align}%
Here $k$ are chosen as discrete values
\begin{equation}
k=\frac{2\pi l}{N},\text{ for }l=1,2,\cdots ,N.
\end{equation}

\begin{figure}[ptb]
\includegraphics[bb=28 408 463 780, width=8 cm, clip]{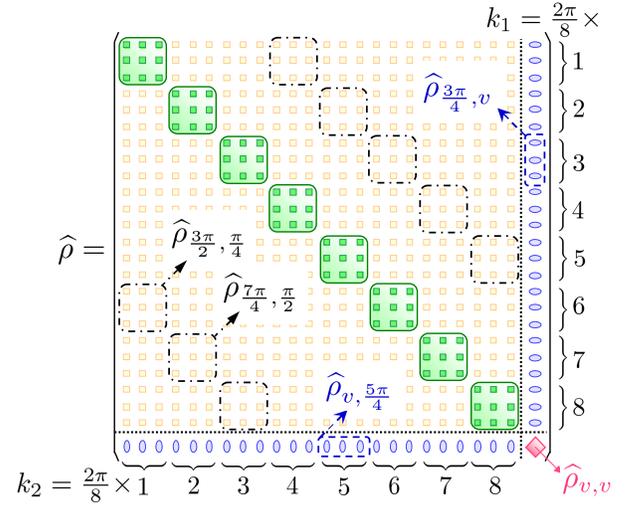}
\caption{(color online) Configuration of the density matrix of the $N=8$
system in the subspace expanded by $\{\left\vert 0\right\rangle , \left\vert
O,k\right\rangle \}$ with $O=A,B,C$ and $k=(2\protect\pi/8)
\times1,2,\cdots,8$. An initial state localized in the $(k_{1},k_{2})$-block
can be evolved to other $(k,k_{2}+k-k_{1})$-blocks (black hollow dot-dash
squares). Only the diagonal $(k,k)$-blocks (green solid squares) are related
to the average transfer time.}
\label{romatrix}
\end{figure}

In the subspace of the single excitation plus the vacuum with the basis%
\begin{equation}
\{\left\vert 0\right\rangle ,\left\vert O,k\right\rangle \equiv O_{k}^{\dag
}\left\vert 0\right\rangle |k=\frac{2\pi l}{N};l=1,2,\cdots N;O=A,B,C\},
\end{equation}%
the general density matrix is decomposed into
\begin{equation}
\widehat{\rho }=\widehat{\rho }_{v,v}+\sum_{k_{1},k_{2}}\widehat{\rho }%
_{k_{1},k_{2}}+\sum_{k}\left( \widehat{\rho }_{v,k}+\widehat{\rho }%
_{k,v}\right) .  \label{rho}
\end{equation}%
where
\begin{equation}
\widehat{\rho }_{v,v}=\rho _{v,v}\left\vert 0\right\rangle \left\langle
0\right\vert
\end{equation}%
is\ the vacuum block while
\begin{equation}
\widehat{\rho }_{k_{1},k_{2}}=\sum_{O,O^{\prime }=A,B,C}\rho
_{Ok_{1},O^{\prime }k_{2}}\left\vert O,k_{1}\right\rangle \left\langle
O^{\prime },k_{2}\right\vert .
\end{equation}%
is called the $\left( k_{1},k_{2}\right) $-block. For fixed $k_{1}$ and $%
k_{2}$, $\rho _{Ok_{1},O^{\prime }k_{2}}$ form a matrix
\begin{equation}
\left(
\begin{array}{ccc}
\rho _{Ak_{1},Ak_{2}} & \rho _{Ak_{1},Bk_{2}} & \rho _{Ak_{1},Ck_{2}} \\
\rho _{Bk_{1},Ak_{2}} & \rho _{Bk_{1},Bk_{2}} & \rho _{Bk_{1},Ck_{2}} \\
\rho _{Ck_{1},Ak_{2}} & \rho _{Ck_{1},Bk_{2}} & \rho _{Ck_{1},Ck_{2}}%
\end{array}%
\right) .
\end{equation}%
The $k$-space representation of the density matrix\ is illustrated in Fig. %
\ref{romatrix} for the $N=8$ system.

In the $k$-space, the master equation (\ref{mastereq}) is reduced to
\begin{align}
& \frac{d\widehat{\rho}_{k_{1},k_{2}}}{dt}=-i\left( H_{k_{1}}\widehat{\rho }%
_{k_{1},k_{2}}-\widehat{\rho}_{k_{1},k_{2}}H_{k_{2}}\right)  \notag \\
& +\sum_{O=B,C}\left\{ \frac{\Gamma^{\prime}}{N}\sum_{k}O_{k}^{\dag}O_{k_{1}}%
\widehat{\rho}_{k_{1},k_{2}}O_{k_{2}}^{\dag}O_{k_{2}+k-k_{1}}\right.  \notag
\\
& \left. -\frac{1}{2}\left( \Gamma+\Gamma^{\prime}\right) \left(
O_{k_{1}}^{\dag}O_{k_{1}}\widehat{\rho}_{k_{1},k_{2}}+\widehat{\rho}%
_{k_{1},k_{2}}O_{k_{2}}^{\dag}O_{k_{2}}\right) \right\}  \label{drok1k2}
\end{align}
for all the $k_{1},k_{2}$,
\begin{align}
\frac{d\widehat{\rho}_{k,v}}{dt} & =-iH_{k}\widehat{\rho}_{k,v}-\frac{1}{2}%
\left( \Gamma+\Gamma^{\prime}\right) \sum_{O=B,C}O_{k}^{\dag}O_{k}\widehat{%
\rho}_{k,v},  \notag \\
\frac{d\widehat{\rho}_{v,k}}{dt} & =i\widehat{\rho}_{v,k}H_{k}-\frac{1}{2}%
\left( \Gamma+\Gamma^{\prime}\right) \sum_{O=B,C}\widehat{\rho}%
_{v,k}O_{k}^{\dag}O_{k}
\end{align}
for all the $k$, and
\begin{equation}
\frac{d\widehat{\rho}_{v,v}}{dt}=\sum_{k}\sum_{O=B,C}\Gamma O_{k}\widehat {%
\rho}_{k,k}O_{k}^{\dag}.  \label{dro00}
\end{equation}
The details of the calculation are shown in Appendix B.

We notice that the equations about $\widehat{\rho }_{k,v}$ and $\widehat{%
\rho }_{v,k}$ are completely decoupled from $\widehat{\rho }_{k_{1},k_{2}}$
and $\widehat{\rho }_{v,v}$. It follows from Eq. (\ref{drok1k2}) that when
no dephasing exists, i.e., $\Gamma ^{\prime }=0$, the $\left(
k_{1},k_{2}\right) $-block $\widehat{\rho }_{k_{1},k_{2}}$ is decoupled with
other $\widehat{\rho }_{k_{1}^{\prime },k_{2}^{\prime }}$ for $\left(
k_{1}^{\prime },k_{2}^{\prime }\right) \neq \left( k_{1},k_{2}\right) $.
Thus $\widehat{\rho }_{k_{1},k_{2}}$ only evolves in the $\left(
k_{1},k_{2}\right) $-block. However, when the dephasing is present ($\Gamma
^{\prime }\neq 0$), the term
\begin{equation}
\sum_{O=B,C}\frac{\Gamma ^{\prime }}{N}\sum_{k}O_{k}^{\dag }O_{k_{1}}%
\widehat{\rho }_{k_{1},k_{2}}O_{k_{2}}^{\dag }O_{k_{2}+k-k_{1}}
\end{equation}%
actually induces the coupling between the $\left( k_{1},k_{2}\right) $-block
and the $\left( k,k_{2}+k-k_{1}\right) $-block. The initial $\widehat{\rho }%
_{k_{1},k_{2}}$ may evolves to $\widehat{\rho }_{k,k_{2}+k-k_{1}}$ as time
goes by. A typical example of $\left( k,k_{2}+k-k_{1}\right) $-blocks are
shown by the black hollow dot-dash squares in Fig. \ref{romatrix}. The
momentum difference $k_{1}-k_{2}$ is conserved during the evolution since%
\begin{equation}
k_{2}-k_{1}=(k_{2}+k-k_{1})-k.
\end{equation}%
In addition, Eq. (\ref{dro00}) means that only the $\left( k,k\right) $%
-blocks of the density matrix result in energy transfer, which are marked by
the $8$ green solid squares in Fig. \ref{romatrix}. All the other $k_{1}\neq
k_{2}$ blocks do not affect the transfer efficiency $\eta \left( t\right) $
and average transfer time $\tau $ at all. Especially, the initial component $%
\widehat{\rho }_{k_{1},k_{2}}$ with $k_{1}\neq k_{2}$ will not influence $%
\eta \left( t\right) $ or $\tau $ at any time $t$ afterwards since
it cannot evolve to the blocks with $k_{1}=k_{2}$. Therefore, only
considering the dynamics of the $\left( k,k\right) $-blocks are
enough for the present purpose.

\section{Transfer efficiency and average transfer time with channel
decomposition}

In this section we use the $k$-space representation of master equation to
calculate the average transfer time and transfer efficiency by the standard
open quantum system method. As a highly organized array of chlorophyll
molecules, the LH2 acts cooperatively to shuttle the energy of photons to
elsewhere when sunlight shines on it. In this sense, we use the density
matrix%
\begin{equation}
\widehat{\rho }\left( 0\right) =\sum_{k_{1},k_{2}}\sum_{O,O^{\prime
}=A,B,C}\rho _{Ok_{1},O^{\prime }k_{2}}\left( 0\right) \left\vert
O,k_{1}\right\rangle \left\langle O^{\prime },k_{2}\right\vert
\label{initstate1}
\end{equation}%
to describe the excitations in the initial state. From the discussions in
the last section, only the $k_{1}=k_{2}=k$ blocks relevant to energy
transfer.\ Therefore, there exists an equivalence class of initial states%
\begin{equation}
\left[ \widehat{\rho }^{\prime }\left( 0\right) \right] =\left\{ \widehat{%
\rho }\text{ }|\text{ }\left\langle O,k\right\vert \widehat{\rho }\left\vert
O^{\prime },k\right\rangle =\rho _{Ok,O^{\prime }k}\left( 0\right) \right\}
\end{equation}%
that results in the same transfer efficiency and average transfer time as
that for $\widehat{\rho }\left( 0\right) $. For further use, a special
density matrix is chosen from the equivalence class $\widehat{\varrho }%
\left( 0\right) \in \left[ \widehat{\rho }^{\prime }\left( 0\right) \right] $%
,%
\begin{eqnarray}
\widehat{\varrho }\left( 0\right) &=&\sum_{k}\sum_{O,O^{\prime }=A,B,C}\rho
_{Ok,O^{\prime }k}\left( 0\right) \left\vert O,k\right\rangle \left\langle
O^{\prime },k\right\vert  \notag \\
&=&\sum_{k}\widehat{\varrho }^{\left[ k\right] }\left( 0\right) ,
\end{eqnarray}%
which satisfies $\left\langle O,k_{1}\right\vert \widehat{\varrho }\left(
0\right) \left\vert O^{\prime },k_{2}\right\rangle =\rho _{Ok,O^{\prime
}k}\left( 0\right) $ for $k_{1}=k_{2}=k$, and $\left\langle
O,k_{1}\right\vert \widehat{\varrho }\left( 0\right) \left\vert O^{\prime
},k_{2}\right\rangle =0$ for $k_{1}\neq k_{2}$. $\widehat{\varrho }\left(
0\right) $ plays an equivalent role for determining the transfer efficiency
and average transfer time. Here,
\begin{equation}
\widehat{\varrho }^{\left[ k\right] }\left( 0\right) =\sum_{O,O^{\prime
}=A,B,C}\rho _{Ok,O^{\prime }k}\left( 0\right) \left\vert O,k\right\rangle
\left\langle O^{\prime },k\right\vert
\end{equation}%
is called as the $k$-channel component of the density matrix. According to
the above observation, we first choose every $\widehat{\varrho }^{\left[ k%
\right] }\left( 0\right) $ as the initial state to obtain the final state $%
\widehat{\varrho }^{\left[ k\right] }\left( t\right) $, which gives the $k$%
-channel transfer efficiency at time $t$,%
\begin{equation}
\eta ^{\left[ k\right] }\left( t\right) =\Gamma \int_{0}^{t}\sum_{k^{\prime
}}\sum_{O=B,C}\varrho _{Ok^{\prime },Ok^{\prime }}^{\left[ k\right] }\left(
t^{\prime }\right) dt^{\prime },  \label{etak}
\end{equation}%
and the $k$-channel average transfer time%
\begin{equation}
\tau ^{\left[ k\right] }=\frac{\Gamma }{\overline{\eta }}\int_{0}^{\infty
}t^{\prime }\sum_{k^{\prime }}\sum_{O=B,C}\varrho _{Ok^{\prime },Ok^{\prime
}}^{\left[ k\right] }\left( t^{\prime }\right) dt^{\prime }.  \label{tauk}
\end{equation}%
Then we prove a general proposition:

\textit{For an arbitrary initial state }$\widehat{\rho }\left( 0\right) $%
\textit{\ (Eq. \ref{initstate1}) of the LH2\ complex, the transfer
efficiency at time }$t$\textit{\ and the average transfer time are the sum
of }$\eta ^{\left[ k\right] }\left( t\right) $\textit{\ and }$\tau ^{\left[ k%
\right] }$\textit{\ over all }$k$\textit{-channels, respectively. }%
\begin{eqnarray}
\eta \left( t\right) &=&\sum_{k}\eta ^{\left[ k\right] }\left( t\right)
\notag \\
\tau &=&\sum_{k}\tau ^{\left[ k\right] }.  \label{etatau1}
\end{eqnarray}

In order to prove the above proposition we notice that the effective initial
state $\widehat{\varrho }\left( 0\right) $ evloves to
\begin{equation}
\widehat{\varrho }\left( t\right) =\sum_{k}\widehat{\varrho }^{\left[ k%
\right] }\left( t\right) .  \label{etatau2}
\end{equation}%
Since the corresponding transfer efficiency and average transfer time of $%
\widehat{\varrho }\left( 0\right) $ are%
\begin{eqnarray}
\eta \left( t\right) &=&\Gamma \int_{0}^{t}\sum_{k^{\prime
}}\sum_{O=B,C}\varrho _{Ok^{\prime },Ok^{\prime }}\left( t^{\prime }\right)
dt^{\prime }  \notag \\
\tau &=&\frac{\Gamma }{\overline{\eta }}\int_{0}^{\infty }t^{\prime
}\sum_{k^{\prime }}\sum_{O=B,C}\varrho _{Ok^{\prime },Ok^{\prime }}\left(
t^{\prime }\right) dt^{\prime },  \label{etatau3}
\end{eqnarray}%
Eq. (\ref{etatau1}) is obtained from Eqs. (\ref{etak}), (\ref{tauk}), (\ref%
{etatau2}), and (\ref{etatau3}). Namely, $\eta \left( t\right) $ and $\tau $
are the sum of $\eta ^{\left[ k\right] }\left( t\right) $ and $\tau ^{\left[
k\right] }$ for different momentum $k$ channels.

The present experimental observations \cite{HuXiChe972} have provided some
potential pathways for light-harvesting. One of them originates from the
excitations on the B800 BChl ring. It shows that the excitations are
transferred to the RC through B800 (LH2) $\rightarrow $ B850 (LH2) $%
\rightarrow $ B850 (another LH2) $\rightarrow \cdots \rightarrow $ B875
(LH1) $\rightarrow $ RC. As to our model, the initial state is specialized as%
\begin{equation}
\widehat{\rho }\left( 0\right) =\sum_{k_{1},k_{2}}\rho
_{Ak_{1},Ak_{2}}\left( 0\right) \left\vert A,k_{1}\right\rangle \left\langle
A,k_{2}\right\vert .  \label{initstate2}
\end{equation}%
Accordingly, the $k$-channel component of the effective initial state $%
\widehat{\varrho }\left( 0\right) $ becomes%
\begin{equation}
\widehat{\varrho }^{\left[ k\right] }\left( 0\right) =\rho _{Ak,Ak}\left(
0\right) \left\vert A,k\right\rangle \left\langle A,k\right\vert =\rho
_{Ak,Ak}\left( 0\right) \widehat{\varrho }^{\left[ A,k\right] }\left(
0\right) .
\end{equation}%
Taking%
\begin{equation}
\widehat{\varrho }^{\left[ A,k\right] }\left( 0\right) =\left\vert
A,k\right\rangle \left\langle A,k\right\vert  \label{initAk}
\end{equation}%
as the initial state, we obtain the transfer efficiency and the average
transfer time
\begin{eqnarray}
\eta ^{\left[ A,k\right] }\left( t\right) &=&\Gamma
\int_{0}^{t}\sum_{k^{\prime }}\sum_{O=B,C}\varrho _{Ok^{\prime },Ok^{\prime
}}^{\left[ A,k\right] }\left( t^{\prime }\right) dt^{\prime }  \notag \\
\tau ^{\left[ A,k\right] } &=&\frac{\Gamma }{\overline{\eta }}%
\int_{0}^{\infty }t^{\prime }\sum_{k^{\prime }}\sum_{O=B,C}\varrho
_{Ok^{\prime },Ok^{\prime }}^{\left[ A,k\right] }\left( t^{\prime }\right)
dt^{\prime }.
\end{eqnarray}%
Hereafter, the superscript $\left[ A,k\right] $ denotes that the initial
state is Eq. (\ref{initAk}). Similar to the above analysis about the
proposition, we present a corollary:

\textit{The transfer efficiency }$\eta \left( t\right) $\textit{\ and
average transfer time }$\tau $\textit{\ of the initial state in Eq. (\ref%
{initstate2}) are the weighted average of }$\eta ^{\left[ A,.k\right]
}\left( t\right) $\textit{\ and }$\tau ^{\left[ A,k\right] }$\textit{,
respectively. \ }%
\begin{eqnarray}
\eta \left( t\right) &=&\sum_{k}\rho _{Ak,Ak}\left( 0\right) \eta ^{\left[
A,.k\right] }\left( t\right)  \notag \\
\tau &=&\sum_{k}\rho _{Ak,Ak}\left( 0\right) \tau ^{\left[ A,k\right] }.
\label{corollary}
\end{eqnarray}%
In the following, we will show the analytical and numerical results of $\eta
^{\left[ A,.k\right] }\left( t\right) $ and $\tau ^{\left[ A,k\right] }$.

First we consider the case without dephasing, i.e., $\Gamma ^{\prime }=0$.
The time evolution from initial state $\widehat{\varrho }^{\left[ A,k\right]
}\left( 0\right) $ only takes place in the $\left( k,k\right) $-block.
According to Eq. (\ref{drok1k2}), the master equation of $\widehat{\rho }%
_{k,k}$
\begin{align}
\frac{d\widehat{\rho }_{k,k}}{dt}& =-i\left( H_{k}\widehat{\rho }_{k,k}-%
\widehat{\rho }_{k,k}H_{k}\right)  \notag \\
& -\frac{\Gamma }{2}\sum_{O=B,C}\left( O_{k}^{\dag }O_{k}\widehat{\rho }%
_{k,k}+\widehat{\rho }_{k,k}O_{k}^{\dag }O_{k}\right) .  \label{drokk}
\end{align}%
gives the average transfer time%
\begin{equation}
\tau ^{\left[ A,k\right] }=\frac{\Gamma }{\overline{\eta }}\int_{0}^{\infty
}t^{\prime }\sum_{O=B,C}\varrho _{Ok,Ok}^{\left[ A,k\right] }\left(
t^{\prime }\right) dt^{\prime }.
\end{equation}

When $k=0$, Eq. (\ref{drokk}) about $\rho _{Ok,O^{\prime }k}$ is rearranged
as a system of differential equations about $v_{j}\left( t\right) $ ($%
j=1,\cdots ,4$) and $v_{5}\left( t\right) =\left[ v_{4}\left( t\right) %
\right] ^{\ast }$:

\begin{align}
v_{1}& =\rho _{Ak,Ak},  \notag \\
v_{2}& =\rho _{Bk,Bk}+\rho _{Ck,Ck},v_{3}=\rho _{Bk,Ck}+\rho _{Ck,Bk},
\notag \\
v_{4}& =\rho _{Ak,Bk}+\rho _{Ak,Ck},v_{5}=\rho _{Bk,Ak}+\rho _{Ck,Ak}.
\end{align}%
It is
\begin{align}
\frac{d}{dt}v_{1}\left( t\right) & =ig_{+}\left[ v_{4}\left( t\right)
-v_{5}\left( t\right) \right] ,  \notag \\
\frac{d}{dt}v_{2}\left( t\right) & =-ig_{+}\left[ v_{4}\left( t\right)
-v_{5}\left( t\right) \right] -\Gamma v_{2}\left( t\right) ,  \notag \\
\frac{d}{dt}v_{3}\left( t\right) & =-ig_{+}\left[ v_{4}\left( t\right)
-v_{5}\left( t\right) \right] -\Gamma v_{3}\left( t\right) ,  \notag \\
\frac{d}{dt}v_{4}\left( t\right) & =2ig_{+}v_{1}\left( t\right) -ig_{+}\left[
v_{2}\left( t\right) +v_{3}\left( t\right) \right]  \notag \\
& -\left[ 2i\left( J_{1}-J_{2}\right) +i\Delta \Omega +\frac{\Gamma }{2}%
\right] v_{4}\left( t\right) ,  \label{vvv}
\end{align}%
with initial conditions%
\begin{equation}
v_{1}\left( 0\right) =1,v_{2}\left( 0\right) =v_{3}\left( 0\right)
=v_{4}\left( 0\right) =v_{5}\left( 0\right) =0.
\end{equation}%
Here $g_{+}=\left( g_{1}+2g_{2}\right) $. Solving the above differential
equations, we obtain%
\begin{align}
\tau ^{\left[ A,k=0\right] }& =\frac{\Gamma }{\overline{\eta }}%
\int_{0}^{\infty }t^{\prime }v_{2}\left( t^{\prime }\right) dt^{\prime }
\notag \\
& =\frac{g_{+}^{2}+\left( J_{1}-J_{2}+\Delta \Omega /2\right) ^{2}+\Gamma
_{0}^{2}/4}{g_{+}^{2}\Gamma _{0}},  \label{tauk0}
\end{align}%
with $\Gamma _{0}=\Gamma /2$, which is independent of the dimerization
parameter $\delta $. Similarly, when $k=\pm \pi $, the average transfer time
of $\widehat{\varrho }^{\left[ A,k\right] }\left( t_{0}\right) $ is%
\begin{equation}
\tau ^{\left[ A,k=\pm \pi \right] }=\frac{g_{-}^{2}+\left( J_{1}+J_{2}\delta
-\Delta \Omega /2\right) ^{2}+\Gamma _{0}^{2}/4}{g_{-}^{2}\Gamma _{0}},
\label{taukpi}
\end{equation}%
where $g_{-}=\left( g_{1}-2g_{2}\right) $. It is a quadratic function with
respect to $\delta $. The optimal parameter $\delta $ with the shortest
transfer time satisfies
\begin{equation*}
\delta _{\mathrm{opt}}^{\left[ A,k=\pm \pi \right] }=\frac{\Delta \Omega
/2-J_{1}}{J_{2}}.
\end{equation*}%
When $g_{1}=2g_{2}$, Eq. (\ref{taukpi}) shows that $\tau ^{\left[ A,k=\pm
\pi \right] }=\infty $, corresponds to $\overline{\eta }=0$, the energy
transfer is prevented at this time.

\begin{figure}[tbp]
\includegraphics[bb=20 200 520 770, width=8 cm, clip]{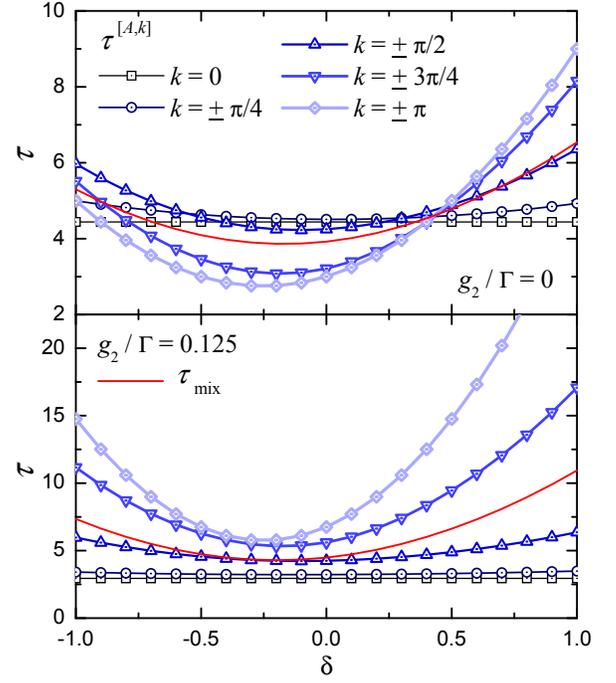}
\caption{(color online) The average transfer time $\protect\tau ^{[A,k]}$ of
$\widehat{\protect\varrho}^{[A,k]}(0) $ (blue scatter lines) and $\protect%
\tau _{\mathrm{mix}}$ of the initial mixed state $\widehat{\protect\rho} _{%
\mathrm{mix}}^{A}(0)$ (red solid lines) with respect to the
dimerization degree $\protect\delta $ of the B850 BChl ring. Here
$N=8$, $J_{1}/\Gamma=0.3$, $J_{2}/\Gamma=1$, $g_{1}/\Gamma=0.5$, $
\Delta\Omega / \Gamma=0.1$, $\Gamma^{\prime }/\Gamma=1$,
$g_{2}/\Gamma=0$ (upper panel) and $g_{2}/\Gamma=0.125$ (lower
panel). $\protect\tau $ is in the
unit of $(1/\Gamma )$ and $\overline{\protect\eta }=1$. It shows that each $%
\protect\tau ^{[A,k]}$ ($k\neq 0$) curve has a minimum at $\protect\delta _{%
\mathrm{opt}}^{\left[ A,k\right] }\neq 0$. $\protect\tau _{\mathrm{mix}}$ is
the equal-weighted average of a complete set of $\left\{ \protect\tau %
^{[A,k]}\right\} $.}
\label{figtau}
\end{figure}

If the dephasing is present, i.e., $\Gamma ^{\prime }\neq 0$, we can provide
approximate solutions for $\tau ^{\left[ A,k=0\right] }$ and $\tau ^{\left[
A,k=\pm \pi \right] }$,%
\begin{eqnarray}
\tau ^{\left[ A,k=0\right] } \!\!&=& \!\! \frac{g_{+}^{2}\!\left( 4\Gamma
_{s} \! -\Gamma ^{\prime }\right) \! \!/\Gamma \! +\!\left( 2J_{1} \!
-2J_{2} \! +\Delta \Omega \right) ^{2}\! \! +\Gamma _{s}^{2}/4}{%
2g_{+}^{2}\Gamma _{s}}  \notag \\
\tau ^{\left[ A,k=\pm \pi \right] } \!\! &=& \!\! \frac{g_{-}^{2}\!\left(
4\Gamma _{s} \! -\Gamma ^{\prime }\right) \!\! /\Gamma \! +\!\left( 2J_{1}
\! +2J_{2}\delta \! -\Delta \Omega \right) ^{2} \!\! +\Gamma _{s}^{2}/4}{%
2g_{-}^{2}\Gamma _{s}},  \label{tauappro}
\end{eqnarray}
where $\Gamma _{s}=\Gamma +\Gamma ^{\prime }$. They almost exactly agree
with the numerical calculation below, and can also be confirmed by Eq. (\ref%
{tauk0}) and (\ref{taukpi}) when $\Gamma ^{\prime }=0$. The details are
shown in Appendix C.

\section{Energy transfer efficiency and average transfer time in numerical calculation}

For a general $k$, the analytical solution of $\eta ^{\left[
A,k\right] }\left( t\right) $ and $\tau ^{\left[ A,k\right] }$ is
not easy to get. Nevertheless, the numerical results of $\tau
^{\left[ A,k\right] }$ as a function of $\delta $ are plotted as
blue scatter lines in Fig. \ref{figtau}. Here we have chosen
\begin{eqnarray}
N &=&8,\frac{J_{1}}{\Gamma }=0.3,\frac{J_{2}}{\Gamma }=1,  \notag \\
\frac{g_{1}}{\Gamma } &=&0.5,\frac{\Delta \Omega }{\Gamma }=0.1,\frac{\Gamma
^{\prime }}{\Gamma }=1,
\end{eqnarray}%
$g_{2}/\Gamma =0$ for the upper panel, $g_{2}/\Gamma =0.125$ for the
lower panel, and $t$ is in the unit of $\left( 1/\Gamma \right) $
and is long enough to ensure $\overline{\eta }=1$. It shows that
when $k\neq 0$
and $\delta $ varies from $-1$ to $1$, there always exist optimum cases $%
\delta _{\mathrm{opt}}^{\left[ A,k\right] }\neq 0$ with and shorter
average transfer time. This fact reflects the enhanced effect of
dimerization.

We then take the mixed initial density matrix $\widehat{\rho }\left(
0\right) =\widehat{\rho }_{\mathrm{mix}}^{A}\left( 0\right) $ as an example,%
\begin{eqnarray}
\widehat{\rho }_{\mathrm{mix}}^{A}\left( 0\right) &=&\sum_{j=1}^{N}\rho
_{Aj,Aj}\left( 0\right) \left\vert A,j\right\rangle \left\langle
A,j\right\vert  \notag \\
&=&\sum_{k_{1},k_{2}}\rho _{Ak_{1},Ak_{2}}\left( 0\right) \left\vert
A,k_{1}\right\rangle \left\langle A,k_{2}\right\vert .
\end{eqnarray}%
The weight $\rho _{Ak,Ak}$ always satisfies%
\begin{equation}
\rho _{Ak,Ak}\left( 0\right) =\frac{1}{N}\sum_{j=1}^{N}\rho _{Aj,Aj}\left(
0\right) =\frac{1}{N}.
\end{equation}%
From Eq. (\ref{corollary}), the transfer efficiency and the average transfer
time of $\widehat{\rho }_{\mathrm{mix}}^{A}$ is%
\begin{eqnarray}
\eta _{\mathrm{mix}}\left( t\right) &=&\frac{1}{N}\sum_{k}\eta
^{\left[
A,k\right] }\left( t\right) \\
\tau _{\mathrm{mix}} &=&\frac{1}{N}\sum_{k}\tau ^{\left[ A,k\right] },
\end{eqnarray}%
$\tau_{\mathrm{mix}}$ is also verified numerically and shown in Fig.
\ref {figtau} as the red solid lines.

\begin{figure}[ptb]
\includegraphics[bb=16 280 530 743, width=8 cm, clip]{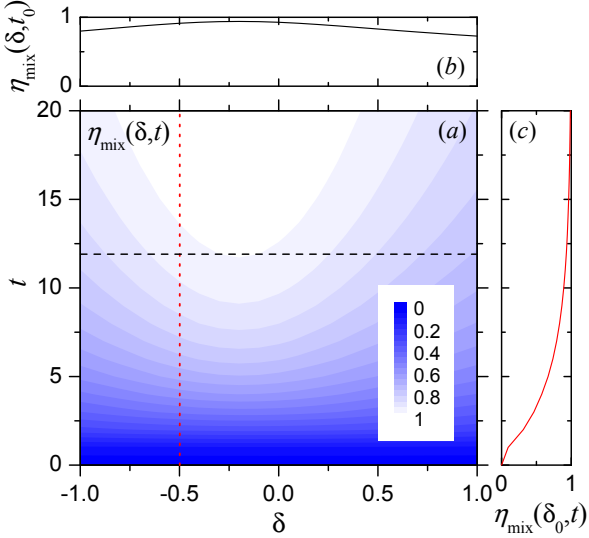}
\caption{(color online) (a) The contour map of the transfer
efficiency of the initial mixed state $\eta
_{\mathrm{mix}}(\delta,t)$ as a function of dimerization degree
$\delta$ and time $t$ for the same setup as that in the lower panel
of Fig. \ref{figtau}. (b) The profile of $\eta _{\mathrm{mix}}$
along $t_{0}=12$ (black dashed line in (a)). (c) The profile of
$\eta _{\mathrm{mix}}$ along $\delta_{0}=-0.5$ (red dot line in
(a)). It shows that $\eta _{\mathrm{mix}}(\delta,t)$ increases over
time, and an optimum $\delta$ can enhance the transfer efficiency.}
\label{figeta}
\end{figure}

In order to see the dynamics of the transfer process clearly, we
plot $\eta _{\mathrm{mix}}$ with respect to the dimerization degree
$\delta$ and time $t$ in Fig. \ref{figeta}(a), i.e., $\eta
_{\mathrm{mix}}=\eta _{\mathrm{mix}}(\delta,t)$. At a certain
instant $t_{0}=12$, $\eta _{\mathrm{mix}}(\delta,t_{0})$ as a
function of $\delta$ is plotted in Fig. \ref{figeta}(b), while for a
certain dimerization degree $\delta_{0}=-0.5$, $\eta
_{\mathrm{mix}}(\delta_{0},t)$ as a function of $t$ is plotted in
Fig. \ref{figeta}(c). Here the parameters are chosen as same as the
ones in Fig. \ref{figtau} except that $g_{2}/\Gamma=0.125$. The
contour map Fig. \ref{figeta}(a) and the profiles of $\eta
_{\mathrm{mix}}(\delta,t)$ in Fig. \ref{figeta}(b) and (c) show that
(1) $\eta _{\mathrm{mix}}(\delta,t)$ increases monotonously as time
goes by. In the large $t$ limit, $\eta _{\mathrm{mix}}(\delta,t)$
equals to $1$. (2) At any certain short instant, an optimum $\delta$
can enhance the transfer efficiency.

Similar to $\eta _{\mathrm{mix}}$ and $\tau _{\mathrm{mix}}$, in
general, there exists an
optimal $\delta _{\mathrm{opt}}\neq 0$ for an arbitrary initial $\widehat{%
\rho }\left( 0\right) $, which means that a suitable distortion of the B850
ring is helpful for the excitation transfer. This result agrees with the
x-ray observation that the Mg-Mg distance between neighboring B850 BChls is
9.2\AA\ within the $\alpha \beta $-heterodimer and 8.9\AA\ between the
heterodimers reported in Ref. \cite{HuXiChe96}. The B850 ring is indeed
dimerized in nature.

As shown in Fig. \ref{figtau}, $\tau ^{\left[ A,k=0\right] }$ and $\tau ^{%
\left[ A,k=\pm \pi \right] }$ are particular since nearly all the other $%
\tau ^{\left[ A,k\right] }$ are within the range of $\left[ \tau ^{\left[
A,k=0\right] },\tau ^{\left[ A,k=\pm \pi \right] }\right] $, so is the
average transfer time $\tau $ of an arbitrary $\widehat{\rho }\left(
0\right) $. Besides, the absolute value of $\delta _{\mathrm{opt}}^{\left[
A,k=\pm \pi \right] }$ for the $k=\pm \pi $ case is larger than the one of
other $\rho \left( 0\right) $, i.e., $\left\vert \delta _{\mathrm{opt}%
}\right\vert \leq \left\vert \delta _{\mathrm{opt}}^{\left[ A,k=\pm \pi %
\right] }\right\vert $.\ Hence, once we have known the properties of $\tau ^{%
\left[ A,k=0\right] }$ and $\tau ^{\left[ A,k=\pm \pi \right] }$, the
behavior of a general $\tau $ can be conjectured to some extend. Compared
the lower panel of Fig. \ref{figtau} with the upper panel, a larger $%
g_{2}/\Gamma $ can increase $\tau ^{\left[ A,k=\pm \pi \right] }$ but
decrease $\tau ^{\left[ A,k=0\right] }$. In the $g_{2}/\Gamma =0.125$ case,
the homogeneous pure state $\widehat{\varrho }^{\left[ A,k=0\right] }\left(
0\right) $ is better than the mixed state $\widehat{\rho }_{\mathrm{mix}%
}^{A}\left( 0\right) $ for energy transport. However, the upper panel with $%
g_{2}/\Gamma =0$ gives the contrary result.

The minimal $\tau ^{\left[ A,k=\pm \pi \right] }$ is reachable at $\delta _{%
\mathrm{opt}}^{\left[ A,k=\pm \pi \right] }=\left( \Delta \Omega
/2-J_{1}\right) /J_{2}$,
\begin{equation}
\tau _{\min }^{\left[ A,k=\pm \pi \right] }=\frac{\left( g_{1}-2g_{2}\right)
^{2}\left( 4\Gamma +3\Gamma ^{\prime }\right) +\Gamma \left( \Gamma +\Gamma
^{\prime }\right) ^{2}/4}{2\left( g_{1}-2g_{2}\right) ^{2}\Gamma \left(
\Gamma +\Gamma ^{\prime }\right) }.
\end{equation}%
In the toy model illustrated in Fig. \ref{figtau}, $J_{2}>J_{1}$ and $%
g_{2}<g_{1}$. When
\begin{equation}
g_{2}/g_{1}=\gamma _{g}=\frac{1}{2}+\xi ^{2}-\xi \sqrt{1+\xi ^{2}},
\end{equation}%
we have $\tau _{\min }^{\left[ A,k=\pm \pi \right] }=\tau ^{\left[ A,k=0%
\right] }$, where
\begin{equation}
\xi =\frac{\left( \Gamma +\Gamma ^{\prime }\right) }{2\left(
2J_{2}-2J_{1}-\Delta \Omega \right) }.
\end{equation}%
On the side of $0<g_{2}/g_{1}<\gamma _{g}$, $\tau _{\min }^{\left[ A,k=\pm
\pi \right] }<\tau ^{\left[ A,k=0\right] }$, while on the\ other side $%
\gamma _{g}<g_{2}/g_{1}<1$, $\tau _{\min }^{\left[ A,k=\pm \pi \right]
}>\tau ^{\left[ A,k=0\right] }$.

\begin{figure}[ptb]
\includegraphics[bb=30 357 550 772, width=8 cm, clip]{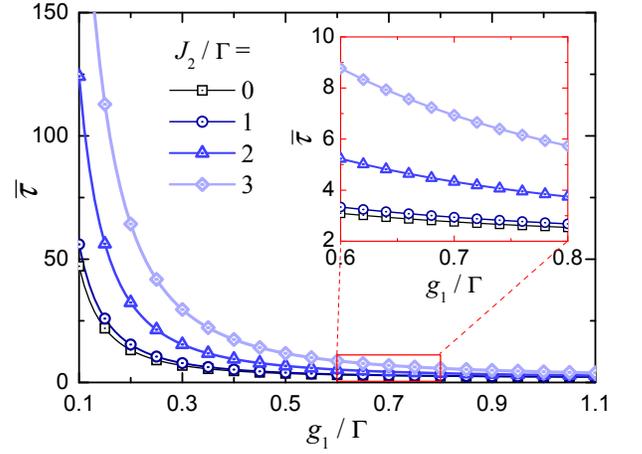}
\caption{(color online) Plots of $\overline{\protect\tau}$ as a function of
the dissipation ratio $g_{1}/\Gamma$ with $J_{2}/\Gamma=0,1,2,3$, where $%
J_{1}/\Gamma=0.3$, $\protect\delta=(\Delta\Omega/2-J_{1})/J_{2}$, $%
g_{2}/g_{1}=0.25$, $\Delta\Omega / \Gamma=0.1$, $\Gamma^{\prime}/\Gamma=0.5$%
, and $\overline{\protect\tau}$ is in the unit of $(1/\Gamma) $. It shows
that $\overline{\protect\tau}$ decreases as $g_{1}/\Gamma$ increases, but
increases with the increasing of $J_{2}/\Gamma$. }
\label{tao2}
\end{figure}

In general, the shortest average transfer time of an arbitrary initial $%
\widehat{\rho }\left( 0\right) $ is within the range of $\left[ \tau _{\min
}^{\left[ A,k=\pm \pi \right] },\tau ^{\left[ A,k=0\right] }\right] $. The
mean value of $\tau _{\min }^{\left[ A,k=\pm \pi \right] }$ and $\tau ^{%
\left[ A,k=0\right] }$ can roughly reflect the influence of parameters on
the transfer process,%
\begin{equation}
\overline{\tau }=\frac{1}{2}\left( \tau _{\min }^{\left[ A,k=\pm \pi \right]
}+\tau ^{\left[ A,k=0\right] }\right) .
\end{equation}%
In Fig. \ref{tao2}, we plot $\overline{\tau }$ with respect to $g_{1}/\Gamma
$ for different $J_{2}/\Gamma =0,1,2,3$. Here, $J_{1}/\Gamma =0.3$, $\delta
=\left( \Delta \Omega /2-J_{1}\right) /J_{2}$, $g_{2}/g_{1}=0.25$, $\Delta
\Omega /\Gamma =0.1$, $\Gamma ^{\prime }/\Gamma =0.5$, and $\overline{\tau }$
is in the unit of $\left( 1/\Gamma \right) $. It shows that $\overline{\tau }
$ decreases monotonously as $g_{1}/\Gamma $ increases. In the short $%
g_{1}/\Gamma $ limit, $\overline{\tau }$ tends to infinity, which is
reasonable since the two BChl rings are decoupled in this case. Moreover, $%
\overline{\tau }$ is larger when $J_{2}/\Gamma $ is larger.

\section{Conclusion}

In summary, we have studied the craggy transfer in light-harvesting
complex with dimerization. We employed the open quantum system
approach to show that the dimerization of the B850 BChl ring can
enhance the transfer efficiency and shorten the average transfer
time for different initial states with various quantum superposition
properties. Actually our present investigation only focuses on a
crucial stage in photosynthesis -- the energy transfer, which is
carried by the coherent excitations in the typical light-harvesting
complex II (LH2). Here the LH2 is modeled as two coupled
bacteriochlorophyll (BChl) rings. With this modeling, the ordinary
photosynthesis is roughly described as three basic steps: 1)
stimulate an excitation in LH2; 2) transfer it to another LH2 or
LH1; 3) the energy causes the chemical reaction that converts carbon
dioxide into organic compounds. Namely, the excitations are
transferred to the RC through B800 (LH2) $\rightarrow $ B850 (LH2)
$\rightarrow $ B850 (another LH2) $\rightarrow \cdots \rightarrow $
B875 (LH1) $\rightarrow $ RC. Obviously, the first two\ are of
physics, thus our present approach can be generalized to investigate
these physical processes. Although photosynthesis happens in
different fashions for different species, some features are always
in common from the point of view of physics. For example, the
photosynthetic process always starts from the light absorbing and
energy transfer.

Another important issues of the photosynthesis physics concerns
about the quantum natures of light \cite{Glauber1,Glauber2}. Since
the experiments have illustrated the role of the quantum coherence
of collective excitations in LH complexes, it is quite natural to
believe that the excitation coherence may be induced by the higher
coherence of photon. Therefore, in a forthcoming paper we will
report our systematical investigation on how the statistical
properties of quantum light affects the photosynthesis.

\acknowledgments

This work is supported by NSFC No. 10474104, 60433050, 10874091 and No.
10704023, NFRPC No. 2006CB921205 and 2005CB724508.

\appendix

\section{Equivalent non-Hermitian Hamiltonian}

In the case without dephasing, i.e., $\Gamma _{j}^{\prime }=0$, an
equivalent non-Hermitian Hamiltonian is introduced to study the dynamics of
the open system,
\begin{equation}
H=H_{S}-i\sum_{j=1}^{N}\frac{\Gamma _{j}}{2}\left( B_{j}^{\dag
}B_{j}+C_{j}^{\dag }C_{j}\right) .  \label{nonHermitian}
\end{equation}%
The equivalence between Eq. (\ref{nonHermitian}) and (\ref{mastereq}) is
shown as follows. On the one hand, the Schr\"{o}dinger equation%
\begin{equation}
i\frac{d}{dt}\left\vert \psi \right\rangle =\left[ H_{S}-i\sum_{j=1}^{N}%
\frac{\Gamma _{j}}{2}\left( B_{j}^{\dag }B_{j}+C_{j}^{\dag }C_{j}\right) %
\right] \left\vert \psi \right\rangle
\end{equation}%
and its Hermitian conjugate%
\begin{equation}
-i\frac{d}{dt}\left\langle \psi \right\vert =\left\langle \psi \right\vert %
\left[ H_{S}+i\sum_{j=1}^{N}\frac{\Gamma _{j}}{2}\left( B_{j}^{\dag
}B_{j}+C_{j}^{\dag }C_{j}\right) \right] ,
\end{equation}%
gives the evolution equation of the density matrix $\widehat{\rho }%
=\left\vert \psi \right\rangle \left\langle \psi \right\vert $,%
\begin{align}
\frac{d\widehat{\rho }}{dt}& =\left( \frac{d}{dt}\left\vert \psi
\right\rangle \right) \left\langle \psi \right\vert +\left\vert \psi
\right\rangle \left( \frac{d}{dt}\left\langle \psi \right\vert \right)
\notag \\
& =-i\left[ H_{S},\widehat{\rho }\right] -\sum_{j=1}^{N}\frac{\Gamma _{j}}{2}%
\left\{ B_{j}^{\dag }B_{j}+C_{j}^{\dag }C_{j},\widehat{\rho }\right\} .
\label{ro1}
\end{align}

On the other hand, when the dephasing terms are absent, the master equation
Eq. (\ref{mastereq}) becomes%
\begin{equation}
\frac{d\widehat{\rho }}{dt}=-i\left[ H_{S},\widehat{\rho }\right]
+\sum_{j=1}^{N}\sum_{O=B,C}\Gamma _{j}[O_{j}\widehat{\rho }O_{j}^{\dag }-%
\frac{1}{2}\left\{ O_{j}^{\dag }O_{j},\widehat{\rho }\right\} ].  \label{ro2}
\end{equation}%
The above equation is written on the expanded Hilbert space with an additive
vacuum basis $\left\vert 0\right\rangle $. Compared with Eq. (\ref{ro1}),
the additive term in Eq. (\ref{ro2}) $\sum_{j=1}^{N}\sum_{O=B,C}$ $\Gamma
_{j}O_{j}\widehat{\rho }O_{j}^{\dag }$ has only contribution to $d\widehat{%
\rho }_{v,v}/dt$, which does not change the dynamics of the system. The Eqs.
(\ref{ro1}) and (\ref{ro2}) are equivalent for determining the time
evolution of $\widehat{\rho }_{Oj,O^{\prime }j^{\prime }}$.

For the non-Hermitian Hamiltonian, the corresponding transfer efficiency and
the average transfer time are%
\begin{eqnarray}
\eta \left( t\right) &=&\int_{0}^{t}\sum_{j=1}^{N}\Gamma
_{j}\sum_{O=B,C}\left\vert \left\langle 0\right\vert O_{j}\left\vert \psi
\left( t^{\prime }\right) \right\rangle \right\vert ^{2}dt^{\prime }  \notag
\\
\tau &=&\frac{1}{\overline{\eta }}\int_{0}^{\infty }t^{\prime
}\sum_{j=1}^{N}\Gamma _{j}\sum_{O=B,C}\left\vert \left\langle 0\right\vert
O_{j}\left\vert \psi \left( t^{\prime }\right) \right\rangle \right\vert
^{2}dt^{\prime }  \label{etataunonH}
\end{eqnarray}%
Due to the equivalence of the non-Hermitian Hamiltonian and the dissipative
master equation, the results of Eq. (\ref{etataunonH}) are as same as the
ones calculated by Eq. Eqs. (\ref{Eeta}) and (\ref{Etau}). The non-Hermitian
Hamiltonian method has an advantage over the master equation one for saving
computer time. Instead of $N^{2}$ equations, only a system of $N$ equations
are needed to be solved in the non-Hermitian Hamiltonian case.

However, when the dephasing terms are present, there is no equivalent
non-Hermitian Hamiltonian. In this case, compared with Eq. (\ref{ro1}), the
additional term $\sum_{O=B,C}\Gamma _{j}^{\prime }$ $O_{j}^{\dag }O_{j}%
\widehat{\rho }O_{j}^{\dag }O_{j}$ cannot be omitted any more. It can also
affect the evolution of the density matrix of the LH2 system.

\section{Transform the master equation to the $k$-space}

In this section, we will transform the master equation from the real space
(Eqs. (\ref{mastereq})-(\ref{dephasing})) to the $k$-space (Eqs. (\ref%
{drok1k2})-(\ref{dro00})). Since $H_{S}=\sum_{k}H_{k}$, and $\widehat{\rho }$
is expressed as Eq. (\ref{rho}), we have%
\begin{eqnarray}
-i\left[ H_{S},\widehat{\rho }\right] &=&-i\left[ \sum_{k}H_{k},%
\sum_{k_{1},k_{2}}\widehat{\rho }_{k_{1},k_{2}}+\sum_{k}\left( \widehat{\rho
}_{v,k}+\widehat{\rho }_{k,v}\right) \right]  \notag \\
&=&-i\sum_{k_{1},k_{2}}\left( H_{k_{1}}\widehat{\rho }_{k_{1},k_{2}}-%
\widehat{\rho }_{k_{1},k_{2}}H_{k_{2}}\right)  \notag \\
&&-i\sum_{k}\left( H_{k}\widehat{\rho }_{k,v}-\widehat{\rho }%
_{v,k}H_{k}\right) .  \label{B1}
\end{eqnarray}%
Here $\left[ H_{S},\widehat{\rho }_{v,v}\right] =0$ since they are in the
different subspaces.

According to the Fourier transformation Eq. (\ref{fourier}), the term $%
\sum_{j}O_{j}\widehat{\rho }O_{j}^{\dag }$ becomes%
\begin{eqnarray}
&&\sum_{j}O_{j}\widehat{\rho }O_{j}^{\dag }=\sum_{k_{1},k_{2},k_{3},k_{4}}%
\frac{1}{N}\sum_{j}e^{i\left( k_{3}-k_{4}\right) j}O_{k_{3}}\widehat{\rho }%
_{k_{1},k_{2}}O_{k_{4}}^{\dag }  \notag \\
&=&\sum_{k_{1},k_{2}}\frac{1}{N}\sum_{j}e^{i\left( k_{1}-k_{2}\right)
j}O_{k_{1}}\widehat{\rho }_{k_{1},k_{2}}O_{k_{2}}^{\dag }  \notag \\
&=&\sum_{k_{1},k_{2}}\delta _{k_{1},k_{2}}O_{k_{1}}\widehat{\rho }%
_{k_{1},k_{2}}O_{k_{2}}^{\dag }=\sum_{k}O_{k}\widehat{\rho }%
_{k,k}O_{k}^{\dag }.
\end{eqnarray}%
The term $\sum_{j}O_{j}^{\dag }O_{j}\widehat{\rho }$ is transformed as%
\begin{eqnarray}
&&\sum_{j}O_{j}^{\dag }O_{j}\widehat{\rho }  \notag \\
&=&\sum_{k_{1},k_{3},k_{4}}\frac{1}{N}\sum_{j}e^{-i\left( k_{3}-k_{4}\right)
j}O_{k_{3}}^{\dag }O_{k_{4}}\left( \sum_{k_{2}}\widehat{\rho }_{k_{1},k_{2}}+%
\widehat{\rho }_{k_{1},v}\right)  \notag \\
&=&\sum_{k_{1},k_{3}}\frac{1}{N}\sum_{j}e^{-i\left( k_{3}-k_{1}\right)
j}O_{k_{3}}^{\dag }O_{k_{1}}\left( \sum_{k_{2}}\widehat{\rho }_{k_{1},k_{2}}+%
\widehat{\rho }_{k_{1},v}\right)  \notag \\
&=&\sum_{k_{1},k_{3}}\delta _{k_{1},k_{3}}O_{k_{3}}^{\dag }O_{k_{1}}\left(
\sum_{k_{2}}\widehat{\rho }_{k_{1},k_{2}}+\widehat{\rho }_{k_{1},v}\right)
\notag \\
&=&\sum_{k_{1},k_{2}}O_{k_{1}}^{\dag }O_{k_{1}}\widehat{\rho }%
_{k_{1},k_{2}}+\sum_{k}O_{k}^{\dag }O_{k}\widehat{\rho }_{k,v}.
\end{eqnarray}%
Similarly,%
\begin{equation}
\sum_{j}\widehat{\rho }O_{j}^{\dag }O_{j}=\sum_{k_{1},k_{2}}\widehat{\rho }%
_{k_{1},k_{2}}O_{k_{2}}^{\dag }O_{k_{2}}+\sum_{k}\widehat{\rho }%
_{v,k}O_{k}^{\dag }O_{k}.
\end{equation}%
Finally, the term $\sum_{j}O_{j}^{\dag }O_{j}\widehat{\rho }O_{j}^{\dag
}O_{j}$ is written as%
\begin{eqnarray}
&&\sum_{j}O_{j}^{\dag }O_{j}\widehat{\rho }O_{j}^{\dag }O_{j}  \notag \\
&=&\sum_{k_{1},k_{2},k_{3},k_{4},k_{5},k_{6}}\frac{1}{N^{2}}%
\sum_{j}e^{-i\left( k_{3}-k_{4}+k_{5}-k_{6}\right) j}O_{k_{3}}^{\dag
}O_{k_{4}}\widehat{\rho }_{k_{1},k_{2}}O_{k_{5}}^{\dag }O_{k_{6}}  \notag \\
&=&\sum_{k_{1},k_{2},k_{3},k_{6}}\frac{1}{N^{2}}\sum_{j}e^{-i\left(
k_{3}-k_{1}+k_{2}-k_{6}\right) j}O_{k_{3}}^{\dag }O_{k_{1}}\widehat{\rho }%
_{k_{1},k_{2}}O_{k_{2}}^{\dag }O_{k_{6}}  \notag
\end{eqnarray}%
\begin{eqnarray}
&=&\frac{1}{N}\sum_{k_{1},k_{2},k_{3},k_{6}}\delta
_{k_{6},k_{2}+k_{3}-k_{1}}O_{k_{3}}^{\dag }O_{k_{1}}\widehat{\rho }%
_{k_{1},k_{2}}O_{k_{2}}^{\dag }O_{k_{6}}  \notag \\
&=&\frac{1}{N}\sum_{k_{1},k_{2},k}O_{k}^{\dag }O_{k_{1}}\widehat{\rho }%
_{k_{1},k_{2}}O_{k_{2}}^{\dag }O_{k_{2}+k-k_{1}}.  \label{B5}
\end{eqnarray}%
Therefore, Eqs. (\ref{drok1k2})-(\ref{dro00}) are obtained by summarizing
Eqs. (\ref{B1})-(\ref{B5}).

\section{Approximative master equations for special cases}

In the cases of $k=0$ and $k=\pm \pi $, we have the approximate master
equation,%
\begin{eqnarray}
\frac{d\widehat{\rho }_{k,k}}{dt} &=&-i\left( H_{k}\widehat{\rho }_{k,k}-%
\widehat{\rho }_{k,k}H_{k}\right) +\sum_{O=B,C}\left\{ \Gamma ^{\prime
}O_{k}^{\dag }O_{k}\widehat{\rho }_{k,k}O_{k}^{\dag }O_{k}\right.  \notag \\
&&\left. -\frac{\Gamma +\Gamma ^{\prime }}{2}\left( O_{k}^{\dag }O_{k}%
\widehat{\rho }_{k,k}+\widehat{\rho }_{k,k}O_{k}^{\dag }O_{k}\right)
\right\} .  \label{drokkgammaD}
\end{eqnarray}%
It is verified numerically that the term $\Gamma ^{\prime }O_{k}^{\dag }O_{k}%
\widehat{\rho }_{k,k}O_{k}^{\dag }O_{k}$ in Eq. (\ref{drokkgammaD}) plays
the same role as $(\Gamma ^{\prime }/N)\sum_{k^{\prime }}O_{k^{\prime
}}^{\dag }O_{k}\widehat{\rho }_{k,k}O_{k}^{\dag }O_{k^{\prime }}$ in Eq. (%
\ref{drok1k2}) for $k=0,\pm \pi $. For the $\left( k=0,k=0\right) $-block,
Eq. (\ref{vvv}) becomes%
\begin{align}
\frac{d}{dt}v_{1}\left( t\right) & =ig_{+}\left[ v_{4}\left( t\right)
-v_{5}\left( t\right) \right] ,  \notag \\
\frac{d}{dt}v_{2}\left( t\right) & =-ig_{+}\left[ v_{4}\left( t\right)
-v_{5}\left( t\right) \right] -\Gamma v_{2}\left( t\right) ,  \notag \\
\frac{d}{dt}v_{3}\left( t\right) & =-ig_{+}\left[ v_{4}\left( t\right)
-v_{5}\left( t\right) \right] -\left( \Gamma +\Gamma ^{\prime }\right)
v_{3}\left( t\right) ,  \notag \\
\frac{d}{dt}v_{4}\left( t\right) & =2ig_{+}v_{1}\left( t\right) -ig_{+}\left[
v_{2}\left( t\right) +v_{3}\left( t\right) \right]  \notag \\
& -\left[ 2i\left( J_{1}-J_{2}\right) +i\Delta \Omega +\frac{\Gamma +\Gamma
^{\prime }}{2}\right] v_{4}\left( t\right) ,
\end{align}%
Solving the above differential equation we have $\tau ^{\left[ A,k=0\right]
} $ shown in Eq. (\ref{tauappro}). The average transfer time $\tau ^{\left[
A,k=\pm \pi \right] }$ for the $k=\pm \pi $ channel is also obtained
similarly.

\end{document}